\documentclass[conference]{IEEEtran}

\usepackage[utf8]{inputenc}
\usepackage{times}
\usepackage{url}
\usepackage{latexsym}
\usepackage{amsmath}
\usepackage{graphicx}
\usepackage{enumerate}
\usepackage{array}
\usepackage[square,sort,comma,numbers]{natbib}
\usepackage{tikz}
\usepackage{verbatim}
\usepackage{subcaption}
\usepackage{multirow}
\usepackage{amssymb}

\usepackage[linesnumbered,ruled]{algorithm2e}
\usepackage[noend]{algpseudocode}
\makeatletter
\def\BState{\State\hskip-\ALG@thistlm}
\makeatother

\usepackage{lipsum} 

\usepackage[hmarginratio=1:1,top=32mm,columnsep=20pt,margin=2cm]{geometry} 

\usepackage{multicol} 
\usepackage{caption} 

\usepackage{booktabs} 
\usepackage{float} 

\usepackage{lettrine} 
\usepackage{paralist} 

\usepackage{abstract} 

\title{Malytics: A Malware Detection Scheme}

\author{\IEEEauthorblockN{Mahmood Yousefi-Azar\IEEEauthorrefmark{1}\IEEEauthorrefmark{2}, Len Hamey\IEEEauthorrefmark{1}, Vijay Varadharajan\IEEEauthorrefmark{3}, Shiping Chen\IEEEauthorrefmark{2}\\~}
\IEEEauthorblockA{\IEEEauthorrefmark{1}Department of Computing, Faculty of Science and Engineering, Macquarie University, Sydney, NSW, Australia.\\
Email: mahmood.yousefiazar@hdr.mq.edu.au, len.hamey@mq.edu.au\\~}
\IEEEauthorblockA{\IEEEauthorrefmark{3}Faculty of Engineering and Built Environment, University of Newcastle.\\
Email: vijay.varadharajan@newcastle.edu.au~}
\IEEEauthorblockA{\IEEEauthorrefmark{2} Commonwealth Scientific and Industrial Research Organisation, CSIRO, Data61.\\
Email: Shiping.Chen@data61.csiro.au~}
}

\begin{document}

\maketitle

\begin{abstract}

An important problem of cyber-security is malware analysis. Besides good precision and recognition rate, a malware detection scheme needs to be able to generalize well for novel malware families (a.k.a zero-day attacks). It is important that the system does not require excessive computation particularly for deployment on the mobile devices.

In this paper, we propose a novel scheme to detect malware which we call Malytics. It is not dependent on any particular tool or operating system. It extracts static features of any given binary file to distinguish malware from benign. Malytics consists of three stages: feature extraction, similarity measurement and classification. The three phases are implemented by a neural network with two hidden layers and an output layer. We show feature extraction, which is performed by \emph{tf}-simhashing, is equivalent to the first layer of a particular neural network. We evaluate Malytics performance on both Android and Windows platforms. Malytics outperforms a wide range of learning-based techniques and also individual state-of-the-art models on both platforms. We also show Malytics is resilient and robust in addressing zero-day malware samples. The F1-score of Malytics is $97.21\%$ and $99.45\%$ on Android dex file and Windows PE files respectively, in the applied datasets. The speed and efficiency of Malytics are also evaluated.

\end{abstract}

\section{Introduction}

Malware detection is of paramount importance to our digital era and thus the daily life. Over 600 millions malware for Windows and 19 million for Android devices were developed in 2016/2017 \cite{AVTest}. In addition to the volume of malware generated, novel families make the detection task overwhelming. 

Malware detection is mostly based on static or/and dynamic analysis of samples \cite{wong2016intellidroid,mariconti_mamadroid,yang2013appintent}. Static analysis uses a binary file and/or disassembled code without running it. It is quite efficient, in most cases, but has problems with heavy obfuscation. Dynamic analysis is a better solution for obfuscated samples because it relies on the run-time behaviour, but it is computationally expensive, and the analysis might not see malicious behaviour during testing. Given features extracted, a classic method to detect malicious codes is to generate a signature for every malware sample. The signature-based methods are only good for detecting known malware.

In particular, it is not difficult to create many polymorphic/metamorphic variants of a given malware sample. The new variants easily evade signature-based defence systems. However, the different variants of the same malware typically exhibit similar malicious patterns. Learning the patterns is the given task of most modern malware detection schemes \cite{rieck2008learning,bayer2009scalable}. 

Deep Convolutional Neural Network (CNN) and other deep learning models have been developed to address a wide range of our daily life phenomenons such as vision, speech and NLP \cite{sabour2017dynamic,lecun2015deep}. The motivations behind them are quite intuitive for the given task and make them state-of-the-art for most cases; however, the proposed scheme of this paper outperforms a wide range of such models. This might be because the scheme is particularly developed for our given task. 

This paper presents a novel learning-based scheme that shows robust ability to detect malware compared with existing state-of-the-art learning-based models and other baselines. The proposed scheme which we call \textit{Malytics} is resilient to zero-day samples. 

We named our model Malytics because the intuition behind the scheme is an analytic solution to detect malware. That is, the learning algorithm comes from a top-down theory with a direct solution rather than learning through samples in an iteration fashion. A wide range of learning algorithms have been developed to learn from input samples \cite{russell2003artificial}. We do not argue the capability of learning models that initiate a hypothesis space (i.e. a model) and adapt this hypothesis space into the training samples. However, we propose to use a learning algorithm that is theoretically related to our proposed feature representation. 

The model is an integrated system in which static features are extracted from a binary file and classified by a neural network. Although deep learning models can be this neural network, it is computationally very expensive to use back-propagation to learn a very large feature space. A common solution for this situation is to use random projection techniques \cite{stokes2017attack}. The projected feature space is then fed to the deep neural network. Random projection with our training algorithm shows quite strong results, supported by a theoretical justification.

Figure \ref{MainScheme} present a high level concept of the Malytics. Inspired from Natural Language Processing (NLP) (see section \ref{FeatureExtraction}), the term-frequency (\emph{tf}) of the given binary file is multiplied by the random projection matrix including 1 and -1. The result is called \emph{tf}-simhashing. This process is linear. 

The representation is fed to the next stage/layer where the similarity indices are obtained as the input for classification. To improve classification, generic non-linear features (e.g. the Gaussian kernel) can be used \cite{scholkopf2002learning}. This can cause poor generalization \cite{bengio2006curse}. A motivation that this paper uses Extreme Learning Machine (ELM) (see section \ref{Scheme}) as the supervised classifier is to address the generalization.

\begin{figure*}[ht]
\centering
\includegraphics[scale=0.3]{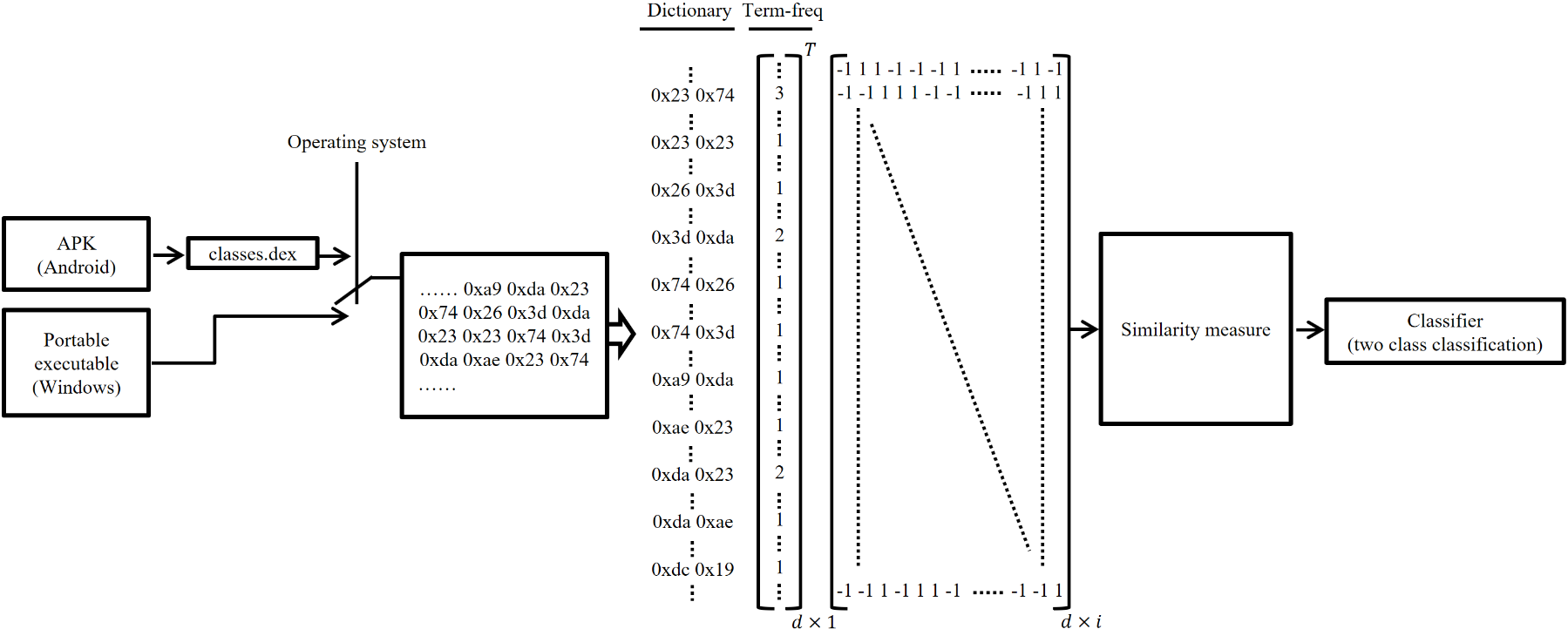}
\caption{Detailed schematic of the proposed solution for malware detection. We used 2-gram (see section \ref{FeatureExtraction}) in this particular example.}
\label{fig:MainScheme}
\end{figure*}

We collected different datasets for our experiments. Because the samples were collected in the wild, they could be a malware file or malicious code that was imaged into another file. This setting helps test the model for real world application. We cannot directly compare our model with the other work because we do not have access to specific state-of-the-art work datasets, except one Android dataset; however, we think that in many cases similar datasets have been used \cite{carlin2017dynamic,wuechner2017leveraging,zhu2017droiddet,raff2017malware}. Our ground-truth for malware samples is a collection of 19 well-known AV vendors.

To evaluate Malytics a wide range of experiments on Android and Windows samples is conducted. For Android, We propose to use the Dex (Dalvik Executable) file rather than the raw APK. Our experiments show \emph{tf}-simhashing of a dex file of an APK carries more efficient information rather than the APK itself. Dex files are also smaller than APKs. The results on Windows Portable Executable (PE) files show that the model is not dependent on a particular operating system. 

Authors \cite{raff2018investigation} showed for n-grams with $n>2$ (e.g 4-gram and 6-gram), the byte-level representation is highly informative while it demands a high computation and feature selection phase. They observed that the information contained in n-grams stem from string features. However, we think 2-gram can carry more information related to instructions and also preserve the string pattern in the frequency. 

{\bf Problem:} Malware must be distinguished from benign samples. The system needs to be fast and efficient. Novel Malware families must be detected. 

{\bf Solution:} Malytics is a resilient solution for the problem. The byte representation contains important information such as APIs, op-codes. The model learns the pattern of bytes. The \emph{tf}-simhashing static feature representation is a fast solution to embed the byte patterns into a short size vector. Malytics generalizes the patterns well even for novel samples. 

{\bf Contribution:}

- We propose a single and integrated model for malware detection. The model has no dependency on particular tools. Malytics places no restriction on the operating system. Evaluation shows it outperforms other single (non-ensemble) state-of-the-art models for both Android and Windows static analysis. 

- We bridge the gap between simhashing and a type of neural network. In particular, we show that simhashing has a close relation to the first layer of ELM. This paper theoretically and empirically show this neural network where the first layer is not trained, has a strong capability for malware detection. 

- We show least squares regression in the form of the ELM with a non-linear kernel can provide a template to fully enhance the feature space rather than implicit feature selection of the regressor used in \cite{raff2018investigation}. The Malytics generalization performance for unseen data also shows the effectiveness of the applied regularization technique.

- A further empirical evaluation of Malytics shows that it can successfully detect new malware families and zero-day samples in the wild. This paper also shows Malytics can be tailored for large scale data application while still remains competitive.

The feature extraction method is presented in section 2, and providing the detail of the proposed scheme in section 3. In section 4, we present a comprehensive evaluation on the performance of Malytics for both Android and Windows platforms, and also discuss the results. In section 5, we provide the limitations of Malytics and future direction of research. Section 5 and 6 present related work to our work and conclusion respectively.

\section{The Feature extraction} \label{FeatureExtraction}

Hashing is a computation which maps arbitrary size data into data of a fixed size. Hashing algorithms have been widely used in the security application domain \cite{jang2011bitshred,dharmapurikar2006fast,manzoor2016fast}. Locality Sensitive Hashing (LSH) is one of the main categories of hashing methods. It hashes input data so that similar data maps to the same “buckets” with high probability, maximizing the probability of a “collision” for similar inputs. Simhashing is one of the most widely used LSH algorithms, adopted to find similar strings \cite{charikar2002similarity,manku2007detecting}. Simhashing is an LSH that is designed to approximate the cosine similarity between inputs. 
The main concept of simhashing comes from Sign Random Projections (SRP) \cite{charikar2002similarity,gionis1999similarity,andoni2006near}. Given an input vector ${V}$, SRP utilizes a random Gaussian unit vector (a random hyper-plane) $I$ with each component generated from a Gaussian unit (i.e., $I_{i} \sim N(0, 1)$ where $i$ is the number of component) and only stores the sign of the projected data as:

\begin{equation}
\textit{hash}({V})=sign({V} \cdot I)
\end{equation}

where $\cdot$ is the dot product. Depending on which side of the hyper-plane ${V}$ lies, $hashing({V})=\pm 1$ . A family of the hash function with the mentioned characteristics provides a setting where for two inputs vectors ${V}$ and ${U}$:

\begin{equation}\label{sim}
\begin{split}
Pr[hash({V})=hash({U})]=1-\frac{\theta({V},{U})}{\pi} \\ \theta=\arccos{(\frac{|{V} \cap {U}|}{\sqrt{|{V}| \cdot |{U}|}})} 
\end{split}
\end{equation}

Where $\theta({V},{U})$ is closely related to $\textit{cosine}({V},{U})$ for the two vectors.  

If $V \cdot I \geq 0$ then $\textit{hash}({V}) = 1$ and otherwise $\textit{hash}({V}) = 0 $, the hamming distance is related to the similarity and it provides a good space to solve the nearest neighbour problem; however, this is not the problem we look to solve. 

The hash function family generates a real value vector if $hash({V}) = {V} \cdot {I}$ and equation \ref{sim} is still guaranteed. We use simhashing that produce real values.

Simhashing has wide-ranging applications from detecting duplicates in texts (e.g. websites) to different security and to malware analysis, specifically with the Hamming distance similarity measure \cite{uddin2011effectiveness,ho2014application,han2014malware}. Inspired from NLP application domain, a n-gram is a contiguous sequence of n items (here, a byte pair) from a given sequence of the binary file. The n-gram feature representation is a specific type of bag-of-words representation in which only the number of occurrences of the items is decisive and the location of the items in the binary file is neglected. The theory behind simhashing allows us to weight the byte n-gram \cite{charikar2002similarity} with the number of occurrences rather than only representing presence (i.e. zero and one) of the byte n-gram in the file.

The proposed feature representation generates a fix size vector from an arbitrary size binary file. Given a binary file, each n-gram is first hashed to a single fix size vector. To speed up this process, first a dictionary of n-grams is provided and then, this vocabulary hashed to binary values. Having \emph{tf} of the vocabulary stored, each hash bit with value 1 or -1 is weighted with \emph{tf} of the n-gram. Thus, \emph{tf} is inserted into the representation \cite{yousefi2017fast}. In the next step, all the vectors sum up bit-wise, thereby providing a final fix size vector. With this process, we embed the distribution of the n-grams of bytes into the vector. This representation provides the two vectors that are close to each other when two files have many common n-grams and different when the files have many different n-grams. 

Bit-wise summing up all the real-value hash vectors of a file (i.e. $hash({V}_{i}) = {V}_{i} \cdot {I}_{i}$ where $i$ is the number of components of hash vector, for example 1024) results in a vector with high variance that needs to be reduced for feeding to an learning algorithm. 

Because we want to map the representation into the space where dot product of vectors directly depends on the angle between vectors, each vector needs to be linearly transformed to have zero mean and unit variance. This transformation is different than normalizing each feature independently to speed up the convergence, because it normalizes each input vector. The other option is to normalize each vector to its Euclidean length (a.k.a $L_2$ norm). In this case, the dot product of two vectors is directly Cosine similarity. But, since the representation will further map to an infinite-dimension space using a Euclidean-distance-based similarity measure, we do not use $L_2$ normalization but linearly transform ($Z = \frac{{X}-\mu}{\sigma}$ to have zero mean and unit variance). We observed that this transformation provides better results. The pseudo-code of the proposed greedy-wise algorithm is:

\begin{algorithm}
\caption{\emph{tf}-simhashing}\label{euclid}
\begin{algorithmic}[1]

{\small
\Procedure{}{} tfSimhash(Dataset, ngram, $i$)
\State $\textit{dictionarySize} \gets 2^{8*\textit{ngram}}$
\State $\textit{componentsSize} \gets \textit{i}$
\State $\bf I_{\textit{dictionarySize} \times \textit{componentsSize}} \gets \text{I}_i \sim {\mathcal{N}(0,1)}$
\State Where $i_i \geq 0$ set to 1 and $i_i<0$ set to -1
\BState \emph{repeat} 
\State \emph{for each binary file}:
\State $\textit{HexStr} \gets Hex(\textit{BinaryFile})$
\State $\textit{TF}_{1 \times dictionarySize} \gets dic(\textit{HexStr}, \textit{dictionarySize})$
\State $\textit{tf-simhash}_{1 \times \textit{componentsSize}} \gets \textit{normalization}(\textit{TF} \times \textit{\bf I})$
\EndProcedure
}
\end{algorithmic}
\end{algorithm}

The aforementioned algorithm is how \emph{tf}-simhashing can be implemented; however, in the context of neural network, algorithm \ref{euclid} is equivalent to the whole process of feeding \emph{tf} representation of byte n-grams to a layer with weights randomly set to 1 or -1; thus, no training is required for this layer (see section \ref{Scheme} for more theoretical elaboration). The output of the hidden layer is exactly our \emph{tf}-simhashing. Indeed, the proposed algorithm bridges the gap between simashing and a neural network in which the first hidden layer has random weights.

We already know that the similarity of the output of the hidden nodes in algorithm \ref{euclid} closely depends on the Cosine angle between two samples (i.e. ${V}$ and ${U}$). In algorithm \ref{euclid}, the vector size $i$ corresponds to the number of hidden nodes in the neural network.

In the next section, we call the \emph{tf}-hashing phase as the first layer of our neural network. 

\section{The Proposed Scheme}\label{Scheme}

Because the latent representation generated in the output of the first hidden layer is based on the similarity of the original space, the second hidden layer of the proposed model can provide a similarity measure. Indeed, we need a task-specific similarity over pairs of data points to facilitate the prior knowledge (i.e. training samples in the first hidden layer). This similarity measure followed by a linear predictor also yields a convex optimization problem \cite{bengio2007scaling}. Kernel methods can play this role. The relation between kernel machines and the neural network has been widely investigated \cite{bengio2007scaling,cho2009kernel,wilson2016deep}. Because kernel layer is data-dependent but unlabeled, the kernel layer training could be seen as unsupervised.

\begin{figure*}[ht]
\centering
\includegraphics[scale=0.25]{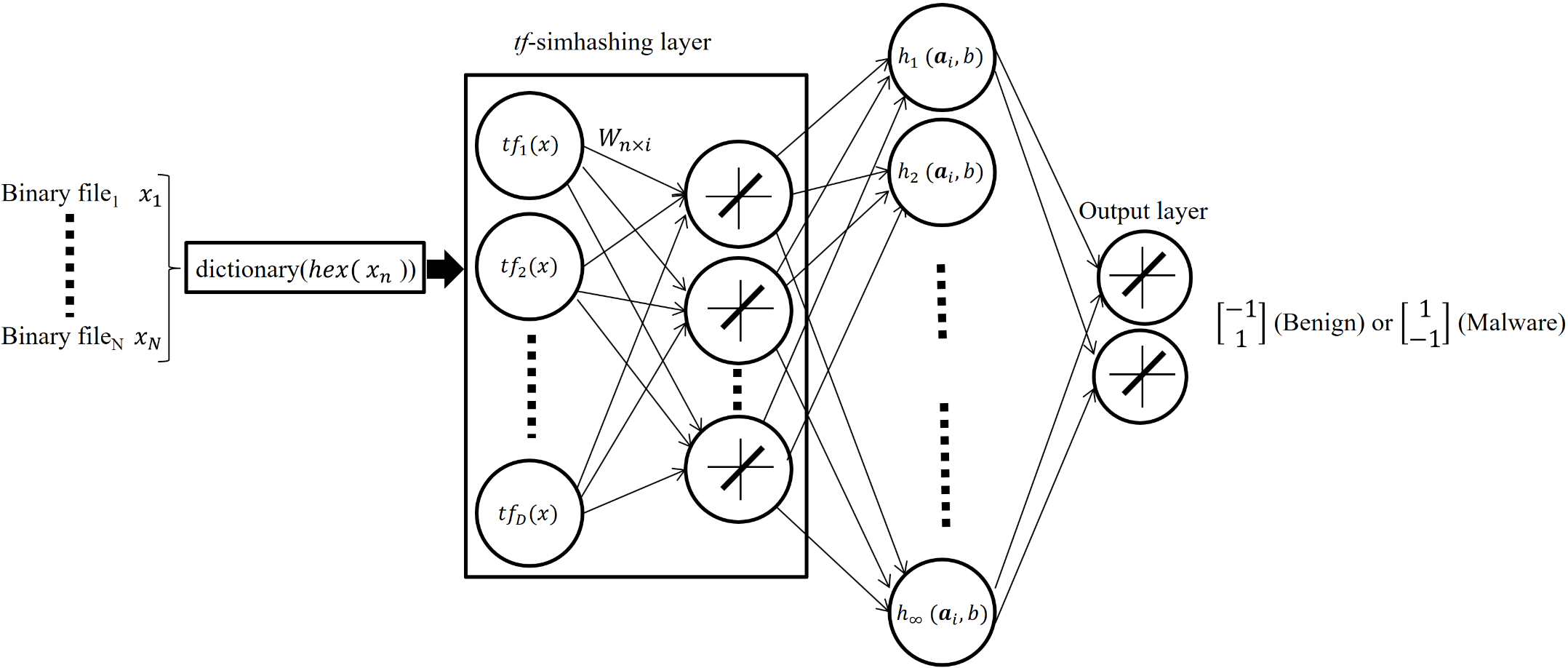}
\caption{The proposed scheme. \emph{tf}-simhashing algorithm has been considered as the first hidden layer.}
\label{fig:ELM}
\end{figure*}

Figure \ref{fig:ELM} presents the proposed scheme. The output layer weights are analytically obtained using the linear least squares technique. The output layer is the ELM. The whole scheme has more than one hidden layer; thus, it is a deep neural network. However, because the training does not use the back-propagation algorithm for training, the scheme differs from the deep learning that is a well-known term in the machine learning community.

The kernel layer is a non-parametric and nonlinear model to match the input to the templates that are obtained from the training samples. The Radial Basis Function (RBF) kernel is well known for providing an infinite-dimensional kernel space and is commonly used with the kernel trick \cite{wilson2016deep,vedaldi2012efficient}.  As we show later, our model supports the kernel trick, so the RBF kernel is a logical choice.

\begin{equation}
    K(x,x_i) = exp^{\left( \frac{ - d( x - x_i )^2 } {2 \gamma ^2 } \right)}
\end{equation}

where $d$ is the Euclidean distance and $\gamma$ is the spread parameter. The function is symmetric $K: \bf X \times \bf X \to \mathbb{R} $, a positive-definite matrix and always a real-valued square matrix. This function projects \emph{tf}-simhashing vectors into an infinite dimensional space. The output layer weight can be trained to predict both classes (i.e. 1 and -1). 

In detail, although approximations to RBF can also provide good results \cite{vedaldi2012efficient}, our malware detection task alongside with the first hidden layer topology give a good reason not to use any approximation but to use the kernel trick. The kernel trick implicitly maps the feature space to an infinite-dimension feature space. The trick makes the mapping limited by the number of data. We empirically evaluate the effect of the kernel dimension (see the subsection \ref{RandomKernel}). \\

To obtain the output layer parameters, we use the ELM. Let $ \{({\bf x}_i,{\bf t}_i)|{\bf x}_i \in \mathbb{R}^d, {\bf t}_i \in \{ -1,1 \} \}_{i=1}^N$, where $N$ is the number of training samples, $d$ is the dictionary size and $m$ is the number of output nodes. The ELM model is $f$ as follows:

\begin{equation}
{\bf f({x})} = \displaystyle\sum_{i=1}^L \beta_i h({\bf x},{\bf{a}}_i,b_i) = \bf h(x) \boldsymbol \beta
\end{equation}

Where $L$ is the number of hidden nodes, $\boldsymbol \beta = [\beta_i, ..., \beta_L]^T$ is the output weights and $\bf a$ and $b$, in our model, are the kernel parameters (i.e $exp^{(- b || {\bf x} - {\bf a}||^2)}$. $b$ are constant when ELM is being trained. $h(\cdot)$ is the RBF kernel. The ELM objective function is to minimize:

\begin{equation}
\begin{split}
&\text{Minimize}_\beta: \frac{1}{2}||\boldsymbol \beta||^2 + C\frac{1}{2} \displaystyle\sum_{i=1}^N ||\xi||^2 \\
&\text{Subject to}: {\bf h}({\bf x}_i)\boldsymbol \beta={\bf t}_{i}^{T} - \xi_{i}^{T}, i=1, ..., N
\end{split}
\end{equation}

Where $C$ is the trade-off parameter, $\xi=[\xi_{i,1}, ..., \xi_{i,m}]^T$ is the error between the desired target (e.g. [-1,1] for benign and [1,-1] for malware) and the labels predicted by the models. There are different techniques to obtain the output layer weights $\boldsymbol \beta$ including orthogonal projection method, iterative methods, and singular value decomposition (SVD) \cite{huang2015trends,golub2012matrix,rao1971generalized}. To minimize the least squares norm, the methods are based on the calculation of the Moore–Penrose pseudo-inverse matrix \cite{penrose1955generalized,rao1971generalized} as follows:

\begin{equation}
\boldsymbol\beta = {\bf H}^\dagger {\bf T}
\end{equation}

Where ${\bf H}^\dagger$ is the matrix. For the sake of feasibility, Kozik \cite{kozik2018distributing} used the SVD matrix factorization technique for malware activity detection; however, the proposed malware detection scheme of this paper allows us to use the closed form solution \cite{huang2014insight}. With Karush-Kuhn-Tucker conditions, the Lagrangian dual problem is defined:

\begin{equation}
\begin{split}
\text{L}_{\text{Dual}_{\text{ELM}}}: \frac{1}{2}||\boldsymbol \beta||^2 + C\frac{1}{2} \displaystyle\sum_{i=1}^N ||\xi||^2 \\ - \displaystyle\sum_{i=1}^N\displaystyle\sum_{j=1}^m \alpha_{i,j}({\bf h}({\bf x}_i)\boldsymbol \beta_j-{\bf t}_{i,j}+\xi_{i,j})
\end{split}
\end{equation}

Where $\boldsymbol \beta_j$ is the out put layer weight/vectors and in our case $m = 2$. The dual problem can be optimized (see Appendix \ref{appendix:B}) and provides the direct solution as follows:

\begin{equation}
\begin{split}
\boldsymbol\beta={\bf H}^T \big(\frac{\bf{I}}{C}+\bf{H}\bf{H}^T)^{-1}\bf{T}
\end{split}
\end{equation}

The ELM function is:

\begin{equation}
\begin{split}
{\bf f(x)} ={\bf h(x)} \boldsymbol \beta = {\bf h(x)} {\bf H}^T \big(\frac{\bf{I}}{C}+\bf{H}\bf{H}^T)^{-1}\bf{T}
\end{split}
\end{equation}

Where $\bf h(\cdot)$ can be unknown and an implicit function satisfies the task. A kernel matrix $\boldsymbol \Omega$ using a kernel function $\bf K$ can be used as follows:

\begin{equation}
\begin{split}
\boldsymbol \Omega = \bf{H}\bf{H}^T: \boldsymbol \Omega_{i,j} &= {\bf h}({\bf x}_i) \cdot {\bf h}({\bf x}_j) \\ &= {\bf K}({\bf x}_i,{\bf x}_j), i,j = 1, ..., N
\end{split}
\end{equation}

The output function as follows:

\begin{equation}
\begin{split}
{\bf f(x)} = \begin{bmatrix}
    {\bf K}({\bf x},{\bf x}_1)\\
    \vdots\\
    {\bf K}({\bf x},{\bf x}_N)
\end{bmatrix} \big(\frac{\bf{I}}{C}+\boldsymbol \Omega)^{-1}\bf{T}
\end{split}
\end{equation}

The applied kernel is RBF. The method is similar to RBF kernel in SVM. Indeed, SVMs are the particular case of ELM. That is, in ELM all the inputs construct support vectors \cite{huang2014insight}. Based on the ELM universal approximation capability that is:

\begin{equation}\label{limitation}
\begin{split}
\lim_{L \to \infty} || \displaystyle\sum_{i=1}^L \boldsymbol\beta_i {\bf h}_i({\bf x})-{\bf f(x)}||=0 
\end{split}
\end{equation}

As long as $\bf h(\cdot)$ is a strict positive definite kernel \cite{deng2016fast,huang2006universal}, a sufficient number of hidden nodes still satisfies \ref{limitation}. In our model, the first hidden layer (\emph{tf}-simashing algorithm) is equivalent to the random nodes; additionally, we also show that we can choose a random subset of the support vectors, from kernel matrix, to reduce the computational over-head and required memory for big data in the cost of the model's performance, that is, $||{\bf K}_{l\times l} \boldsymbol \beta - T|| < \epsilon$ where $l<L$. 

The proposed scheme is summarized in Algorithm \ref{SchemeAlgori}

\begin{algorithm}
    \SetKwInOut{Input}{Input}
    \SetKwInOut{Output}{Output}
    \SetKwInOut{Training}{Training}
    \SetKwInOut{Testing}{Testing}

{\small    

\Input{given $N$ training samples as $\{({\bf x}_i,{\bf t}_i)|{\bf x}_i \in \mathbb{R}^d, {\bf t}_i \in \{ -1,1 \} \}_{i=1}^N$, given $V$ testing samples as $\{({\bf x}_i,{\bf t}_i)|{\bf x}_i \in \mathbb{R}^d, \}_{i=1}^V$, n-gram, $i$, $l$}
    \Output{predicted labels (Benign or Malware)}
    \Training {\\
    $\textit{tf\mbox{-}simhash}_{N \times i} \gets \textit{tfSimhash(Dataset,ngram,i)}$\\
    $KernelMatrix_{N,N} \gets \textit{RBF({tf\mbox{-}simhash}}_{N \times i}$)\\
    \If{$l \ne N$}
      {
        $RandK \gets Random(1:l)$\\
        return $\textit{KernelMatrix}_{l,l}(RandK,RandK)$\;
      }
      
$\boldsymbol \beta \gets \big({\bf\frac{I}{C}}+KernelMatrix \big)^{-1}\bf{T}$ 
}

    \Testing {\\
    $\textit{tf\mbox{-}simhash}_{V \times i} \gets \textit{tfSimhash(binary file,ngram,i)}$\\
    $\textit{Predictions} \gets \boldsymbol \beta \times RBF(\textit{tf\mbox{-}simhash}_{V \times i}$)\\
    }

}   

    \caption{the proposed scheme algorithm}\label{SchemeAlgori}
\end{algorithm}

\section{Evaluation}

To evaluate our scheme, we conducted a wide range of experiments on real datasets collected from the wild for both Android and Windows platforms. During all experiments, we kept the training data balanced (i.e. malware to benign ratio (MBR) is 0.5) except especially to evaluate the capability of the scheme to deal with imbalanced data. Our benign samples were collected from Androzoo, a freely available APks repository \cite{allix2016androzoo}. Androzoo crawled several markets including the official Google Play. We randomly selected apps collected from Google Play. Our collected malware samples (from Virussahre.com\footnote{\url{https://virusshare.com/}}) includes a wide range of malware families for both Android and Microsoft Windows platforms. We double checked the status of all malware samples using VirusTotal.com API\footnote{\url{https://www.virustotal.com/##/home/search}}. VirusTotal provides the results of analysis by about 55 anti-virus vendors. To avoid having a considerable inconsistency, we selected 19 of the most well-known vendors' results. The selected companies are Kaspersky, Symantec, ESET-NOD32, Avast, McAfee, AVG, Avira, Microsoft, BitDefender, Panda, F-Secure, Malwarebytes, TrendMicro, Comodo, VIPRE, AVware, Ad-Aware, Sophos and Qihoo-360. A malware that is detected by at least one of the vendors was picked to be included in our datasets. This procedure was used to establish the ground-truth in all our experiments.

The first malware dataset is Drebin \cite{arp2014drebin}. It consists of 5560 malware of which 5555 have a .dex file. 

\begin{figure}[ht]
\centering
\includegraphics[scale=0.2]{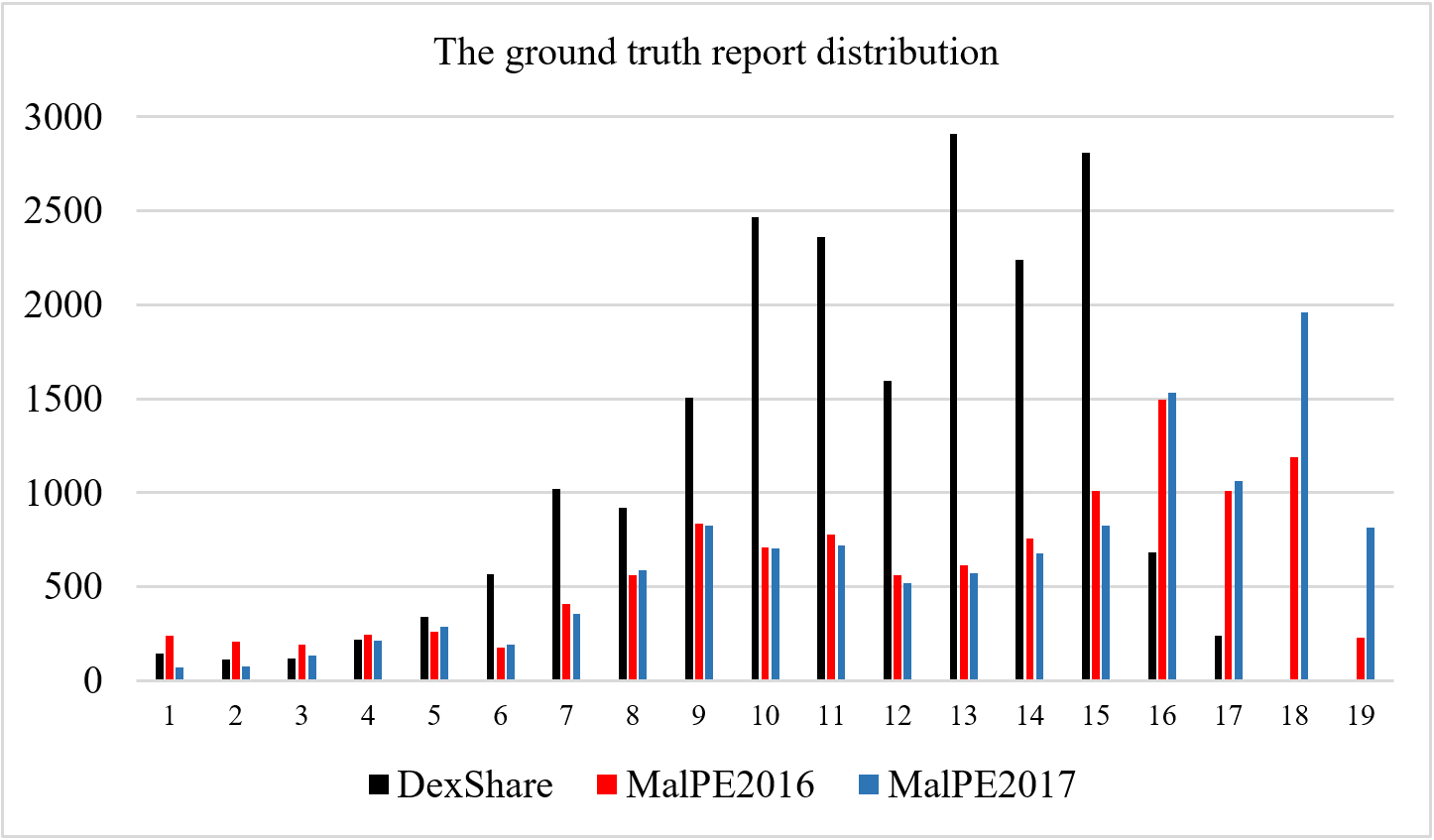}
\caption{The number of samples detected by the 19 selected anti-virus vendors.}
\label{fig:vendorDetection}
\end{figure}

We also collected our own malware dataset from the two packages VirusShare{\_}Android{\_}20130506.zip and VirusShare{\_}Android{\_}20140324.zip, downloaded from VirusShare.com. Together these consist of 35397 malware that were collected before April 2014. Since some samples of the packages have been reversed engineered and were re-compressed, we focus on intact samples. Also, each malware must meet our ground-truth threshold. 20255 malware samples met our criterion. Figure \ref{fig:vendorDetection} shows the number of samples detected by the 19 anti-virus vendors. We again randomly select 20255 benign from our repository. The statistics of the dataset (DexShare) is presented in the table \ref{tab:dataset}.

For Microsoft Windows, 8912 PE (WinPE) benign has been collected from a fresh installed Windows 10 with 2016 updates. The other benign set, consisting of 11983 PE (WinAppPE), was collected by combining the Windows benign with 77 applications (e.g. firefox, Adobe Reader, etc), automatically install by Ninite \footnote{\url{https://ninite.com/}}. For the malware set, we downloaded VirusShare{\_}00271.zip a package containing 65,536 malware of which 11483 are PEs that also meet the threshold of the selected anti-virus vendors. The package was captured from 2016-11-01 to 2016-11-20 from the wild. To provide balance, we collected further 500 malware from the previous package in VirusShare and added this malware set to yield MalPE2016. Because one of our evaluation goals is zero-day detection, we also collected PEs of VirusShare{\_}00298.zip (MalPE2017) that is a package collected about one year after PE2016. The table \ref{tab:dataset} shows the statistics of the Windows dataset (PEShare). Figure \ref{fig:vendorDetection} also shows the number of samples detected by 1 to 19 anti-virus vendors for both malware sets. When it comes to testing Malytics for WinPE set, we randomly select 8912 malware from MalPE2016.

\begin{table}[ht]
\centering
\begin{tabular}{l*{4}{c}c}
\toprule
 \bf Dataset & \bf Type &\bf Qty & \bf Max. &\bf Min. &\bf Ave. \\
 \midrule
 \bf Drebin & Malware & 5555 & 06.0 & 2.3 & 0.36\\  
 & Benign & 5555 & 13.5 & 1.4 & 3.00\\ 
 \midrule
 \bf DexShare & Malware & 20255 & 09.2 & 1.0 & 0.65\\  
 & Benign & 20255 & 10.5 & 1.9 & 1.56\\ 
 \midrule
  \bf PEShare & MalPE2016 & 11983 & 53.0 & 0.8 & 0.97\\  
   & MalPE2017 & 12127 & 54.1 & 1.5 & 1.26\\  
   & WinPE & 8912 & 33.6 & 1.7 & 0.35\\  
 &  WinAppPE & 11983 & 80.3 & 1.5 & 0.50\\ 
 \bottomrule
\end{tabular}
\caption{The statistics of the datasets. Max, Min and Ave stand for maximum size (MB), minimum size (KB) and average size (MB) of the files respectively}
\label{tab:dataset}
\end{table}

We used different evaluation metrics to analyze the performance of the proposed scheme. The metric for our two class classification task is based on the confusion matrix: 

\footnotesize
\vbox{
\centering
\begin{tabular}{c|l|c|c|c}
\multicolumn{4}{c}{} \\
\multicolumn{2}{c}{}&\multicolumn{2}{|c|}{Predicted}&\\
\cline{3-4}
\multicolumn{2}{c|}{}&Benign&Malware\\
\cline{1-4}
\multirow {2}*{Actual}& Benign & True Negative (TN) & False Positive (FP) \\
\cline{2-4}
& Malware & False Negative (FN) & True Positive (TP) \\ \cline{1-4} \multicolumn{4}{c}{}
\end{tabular}
\label{tab:confusionmatrix}
}
\normalsize

The common performance metrics are:

Recall (a.k.a  hit/detection rate or sensitivity) $= \frac{TP}{TP+FN}$. And, $\textit{False Negative Rate (FNR)} = \frac{FN}{TP+FN}$. where FNR is 1 - recall.

$\textit{Precision} = \frac{TP}{TP+FP}$

$\textit{f1-score} = 2*\frac{precision*recall}{precision+recall}$

$\textit{Accuracy} = \frac{TN+TP}{TN+FP+FN+TP}$

$\textit{False positive rate (FPR)} = \frac{FP}{TN+FP}$

The mentioned metrics are enough to evaluate a model. We also used AUC (the area under receiver operating characteristic (ROC) curve) where we found other work with this index. AUC is the probability that a classifier will rank a randomly chosen positive sample higher than a randomly chosen negative sample. 

\subsection{Experimental Setup}

To practically show the scheme's efficiency and meeting the mentioned classifier characteristics, we used Support Vector Machines (SVMs), Gradient Boost (XGBoost), Deep Neural Network (DNN), Random Forest (RF) and K-Nearest Neighbors (K-NN) as our baselines. The scikit-learn Python library was used to implement the baselines for SVM, XGBoost, RF, K-NN. We use Keras API , that runs on top of TensorFlow, to implement Deep Neural network. We used grid search to optimize the hype-parameters of the baselines. The results of this the grid search (range from $10^{-4}$ to $10^4$ for both C and gamma) for SVM $C=0.1$ and $\gamma=100$ with RBF kernel. The grid search for K-NN ranged from 1 to 20 for n\_neighbors and is either ’uniform’ or ’distance’ for weights. The Deep learning model has 3 hidden layer with 1024, 128 and 64 nodes and 1 node in the output layer. The activation functions are linear rectifier, linear rectifier, sigmoid. The first hidden layer is the proposed representation. The optimizer, batch size and the number of epochs are adam, 5, 100. The first two hidden layer adapt dropout regularization with 0.2 probability.

The two hyper-parameters of Malytics are trade-off parameter C and kernel parameter $\gamma$. The result of the grid search for hyper-parameters C (ranging from 10 to 500) is 200 and for $\gamma$ (ranging from 0.5 to 1.5.) is 1. 

Our machine specification is Intel(R) Core(TM) i7-4790 CPU @ 3.60GHz, 32.0 GB RAM and hard disk drive (HDD).

The hashing algorithm was designed to generate the vector of size 1024, showing strong performance. We observe that larger vectors provide the slightly better results and smaller vectors reduce the performance. The vector of size 1204 is an optimum size while it still is computationally cheap. Over our experiments, we choose 2-gram and used 5-fold cross-validation. Malytics even with 2-gram outperforms the state-of-the-art models. We do not fix a threshold to ensure the highest precision and recall but rather the range of the capability of the model is also a goal of this paper.

\subsection{Results and Discussion}

This part of the paper presents the results of the proposed model. To compare the scheme with other models, we feed \emph{tf}-simhashing to different classifiers. Thus, the proposed feature extraction technique is also examined using different classifiers. Because the presented results on both Android and Windows are based on the 5-fold cross-validation, we present mean and standard deviation (Std) of all 5-folds in the tables. It is usual to present FPR as a fix value rather than mean ($\pm \textit{Std}$). We also calculated FPR in this way.

\subsection{Android Malware Detection}

This section presents the performance of Malytics on Android malware detection. First, we show the model's capability compared with different baselines on Drebin and DexShare datasets. Then, the model is evaluated in different settings. Further analysis of Android malware detection is presented in the section \ref{FurtherAnalysisAndCaseStudy}.

Table \ref{tab:mainTable1} shows the performance of Malytics compared with the state-of-the-art models \cite{grosse2017adversarial,mariconti_mamadroid}, 5 baselines and ELM (without kernel layer) on the Drebin dataset. Grosse \cite{grosse2017adversarial} with MBR=0.5 can be compared with our experiments. Mariconti \cite{mariconti_mamadroid} also provides a similar setting to ours. Malytics outperforms all other techniques when it comes to detecting malware, that is, $\textit{FNR} = 1.44\%$. This superior performance is seen for f-score and accuracy as well. Interestingly, DNN is the most precise model compared with all others with only $1\%$ FPR while its FNR is the worst.

The Drebin dataset is a publicly available dataset but not very sophisticated. We also tested the models on a more sophisticated dataset, DexShare, with more samples, collected in wider time windows. We used AVCLASS tool to label malware samples of both Drebin and DexShare \cite{sebastian2016avclass}. The tool labels the malware sets based on VirusTotal reports. Because we use VirusTotal to double-check the collected malware set, reports were available to use AVCLASS for labeling. With the tool, Drebin has 180 malware families while DexShare has 309 families. 

The performance of the models on DexShare is presented in table \ref{tab:mainTable2}. It is to be expected that all models perform weaker on the dataset compared to Drebin, since the detaset is more complicated to deal with. Malytics again outperforms all models on DexShare. The results show that Malytics has the highest hit rate (a.k.a recall = 1-FNR) to detect malware and the highest precision that corresponds to low FPR. 

In addition to Malytics, most baselines also provide good performance compared with Zhu \cite{zhu2017droiddet}. This trend shows the \emph{tf}-simhashing feature representation is rich and many classifiers can leverage it and provide good performance. It is true that Hui-Juan \cite{zhu2017droiddet} did not exactly use the DexShare dataset, but they collected their dataset from Virusshare.com as we did. Virusshare.com has two packages for Android malware, and DexShare is a combination of both. So, the results can be compared. Additionally, Hui-Juan's \cite{zhu2017droiddet} feature extraction is on the basis of static analysis. For example, \emph{tf}-simhashing feeding to SVM yields $93.35\%$ ($\pm 0.16\%$), $08.00\%$ ($\pm 0.48\%$), $94.77\%$ ($\pm 0.25\%$), $93.44\%$ ($\pm 0.16\%$) for AUC, FNR, precision, and accuracy respectively while Hui-Juan \cite{zhu2017droiddet} reported $86.00\%$ ($\pm 2.0\%$), $13.82\%$ ($\pm 2.3\%$), $84.13\%$ ($\pm 3.5\%$) and $84.93\%$ ($\pm 1.8\%$) for AUC, FNR, precision, and accuracy respectively when they used SVM as the classifier.

\begin{table}[ht]
\centering
\scalebox{0.8}{
\begin{tabular}{l*{5}{c}}
\toprule
 \bf Model  & \bf FNR &\bf Precision &\bf f1-score & \bf Accuracy & \bf FPR\\
 \midrule
SVM   & $04.81\%$ &  $96.62\%$ & $95.90\%$ & $95.93\%$ & $ 3.33\%$\\ 
      & $(\pm 1.15\%)$ & $(\pm 0.40\%)$ & $(\pm 0.80\%)$ & $(\pm 0.77\%)$ &\\ 
XGBoost  & $02.97\%$ & $93.59\%$ & $95.28\%$ & $95.19\%$ & $6.64\%$\\ 
   & $(\pm 0.20\%)$ &  $(\pm 0.31\%)$  &  $(\pm 0.23\%)$ & $(\pm 0.25\%)$ &\\ 
DNN & $13.50\%$ &  $ \bf 98.77\%$  & $92.22\%$ & $92.70\%$ & $ \bf 1.00\%$\\ 
 & $(\pm 0.64\%)$ &  $(\pm 0.62\%)$  & $(\pm 0.46\%)$ & $(\pm 0.42\%)$  &\\ 
RF & $04.18\%$ &  $92.52\%$ & $94.14\%$ & $94.03\%$ & $7.76\%$\\ 
 & $(\pm 0.55\%)$ & $(\pm 0.98\%)$  & $(\pm 0.34\%)$ & $(\pm 0.38\%)$&\\
K-NN & $02.80\%$ &  $93.36\%$ &  $95.25\%$ & $95.15\%$ & $6.91\%$\\ 
 & $(\pm 0.48\%)$ &  $(\pm 0.70\%)$ & $(\pm 0.55\%)$ & $(\pm 0.57\%)$ & \\ 
ELM & $ 03.00\%$ &  $94.51\%$ & $ 95.76\%$ & $ 95.70\%$ & $5.60\%$\\ 
 & $(\pm 0.30\%)$ & $(\pm 0.47\%)$ & $(\pm 0.31\%)$ & $(\pm 0.30\%)$  & \\ 
Malytics & $ \bf 01.44\%$ & $ 96.45\%$  & $\bf 97.36\%$ & $ \bf 97.33\%$ & $3.90\%$\\ 
 & $(\pm 0.33\%)$ &  $(\pm 0.45\%)$  & $(\pm 0.29\%)$ & $(\pm 0.30\%)$   &\\ 
 \cite{grosse2017adversarial}  & $06.37\%$ & $-$ & $-$ & $95.93\%$ & $3.96\%$\\ 
 \cite{mariconti_mamadroid} & $ 3.00\%$ & $95.00\%$ & $ 96.00\% $ &   $ - $ & $ - $\\ 
 \bottomrule
\end{tabular}
}
\caption{The Mean and Std of Malytics and the baselines for Drebin Dataset.}
\label{tab:mainTable1}
\end{table}

\begin{table}[ht]
\centering
\scalebox{0.8}{
\begin{tabular}{l*{5}{c}}
\toprule
 \bf Model  & \bf FNR &\bf Precision &\bf f1-score & \bf Accuracy & \bf FPR\\
 \midrule
SVM   & $08.00\%$ & $94.77\%$ & $93.34\%$ & $93.44\%$ & $05.07\%$\\ 
      & $(\pm 0.48\%)$ & $(\pm 0.25\%)$ & $(\pm 0.18\%)$ & $(\pm 0.16\%)$ &\\ 
XGBoost  & $10.12\%$ & $90.74\%$ & $90.30\%$ & $90.35\%$ & $09.17\%$\\ 
   & $(\pm 0.55\%)$ & $(\pm 0.39\%)$ & $(\pm 0.36\%)$ & $(\pm 0.35\%)$&\\ 
DNN & $24.40\%$ & $90.40\%$ & $82.23\%$ & $83.72\%$ & $08.13\%$\\ 
 & $(\pm 3.53\%)$ & $(\pm 1.73\%)$ & $(\pm 1.45\%)$ & $(\pm 0.89\%)$ &\\
RF & $13.07\%$ & $92.73\%$  &  $89.73\%$ & $90.05\%$ & $06.82\%$\\ 
 & $(\pm 0.35\%)$ & $(\pm 0.32\%)$ & $(\pm 0.22\%)$ & $(\pm 0.21\%)$&\\ 
K-NN & $7.45\%$ & $93.36\%$ & $91.38\%$ & $91.28\%$ & $10.00\%$\\ 
 & $(\pm 0.48\%)$ & $(\pm 0.70\%)$ & $(\pm 0.55\%)$ & $(\pm 0.57\%)$ &\\
ELM & $ 16.50\%$ & $ 92.25\%$ & $ 87.66\%$ & $ 88.24\%$ & $ 07.00\%$\\ 
 & $(\pm 0.29\%)$ & $(\pm 0.67\%)$ & $(\pm 0.30\%)$ & $(\pm 0.36\%)$ &\\
Malytics & $ \bf 05.53\%$ & $ \bf 95.88\%$ & $\bf 95.17\%$ & $ \bf 95.20\%$ & $\bf 04.06\%$\\ 
 & $(\pm 0.46\%)$ & $(\pm 0.40\%)$ & $(\pm 0.20\%)$ & $(\pm 0.20\%)$ &\\
  \midrule
 \cite{zhu2017droiddet} & $11.60\%$ & $88.16\%$ & $ - $ &   $88.26\%$ & $ - $\\ 
 & $(\pm 2.76\%)$ & $(\pm 1.8\%)$ & $ - $ & $(\pm 1.73\%)$ &\\ 
 \bottomrule
\end{tabular}
}
\caption{The Mean and Std of Malytics and the baselines for DexShare Dataset.}
\label{tab:mainTable2}
\end{table}

Table \ref{tab:mainTable3} provides more inside into Malytics. The \emph{tf}-simhashing feature extraction algorithm can be implemented on APK as well as only Dex file of respective APK. Yousefi-Azar \cite{yousefi2017fast} is based on \emph{tf}-simhashing of the APK. Table \ref{tab:mainTable3} shows hashing the dex file yields much better performance compared with hashing the whole APKs.

A common test is to evaluate a model in an imbalanced setting to mimic the real world settings. For the test, MBR is typically $10\%$, $20\%$, $30\%$. To have enough malware to test the scheme and also to provide the imbalanced settings, we chose MBR$=0.2$. That is, we randomly selected 5060 malware from the DexShare malware set while the total benign set was used. Table \ref{tab:mainTable3} shows that Malytics performs more precisely with imbalanced data. We expect this results because the model saw more benign sample in the training. FNR does not show a statistically significant change. It demonstrates that Malytics is robust to the imbalanced situations.

One of the most important tests of a malware detection system is to evaluate the system against zero-day malware. There are different evaluation methods to do a zero-day experiment. Mariconti \cite{mariconti_mamadroid} use a time frame test. That is, they trained the model with samples of a given date, the model was tested on samples of one year and also two years later than the given date. In short, training on past sample and testing on new samples in time. 

Although we could use the timestamp of the samples of DexShare, because the timestamp of a file is easily forged, both by malware and benign writers, we think timestamp is not a good index to partition our dataset into past and future samples. We also think that AVCLASS is not a very accurate technique to partition our dataset. 

However, because we do not have any other concrete option, we again rely on AVCLASS. As mentioned earlier in this section, AVCLASS labels DexShare with 309 different families. From 309 families, about 20 families have more than about 150 samples in each family. We chose these 20 families for our novelty detection test. To do this test, we selected training and test sets and we do not use cross-validation. From malware set, out of 20 families, 4 families were chosen to be the test set and the rest of malware set were chosen to be training set. We did this test 5 time to test on all 20 families. To be clear, when 4 families were chosen to be the test set, the other 16 families plus the rest of the malware set are the training set. The benign set was randomly chosen from DexShare to keep the training and test sets balanced. 

Table \ref{tab:mainTable3} shows the average FNR, precision, f1-score, accuracy and FPR of Malytics with our proposed feature representation. The main important index of the test is FNR as a measure that shows how well Malytics detected new families. Mariconti \cite{mariconti_mamadroid} also provided a novelty detection setting based on detecting future malware. Our test is different with their test. But if we assume that our family exclusion test is at least as difficult as predicting future malware (e.g. test on one year in future), we can see that Malytics is quite competitive with the state-of-the-art in novelty detection. For further explanation see section \ref{AndroidFamilyDetectionCase}.

\begin{table}[ht]
\centering
\scalebox{0.75}{
\begin{tabular}{l*{5}{c}}
\toprule
 \bf Model  & \bf FNR &\bf Precision &\bf f1-score & \bf Accuracy & \bf FPR\\
 \midrule
Malytics (APK) & $09.43\%$ & $91.76\%$ & $91.16\%$ & $91.22\%$ & $8.1\%$\\ 
      & $(\pm 0.61\%)$ & $(\pm 0.73\%)$ & $(\pm 0.54\%)$ & $(\pm 0.48\%)$ &\\ 
Malytics (Dex, MBR=0.5) & $05.33\%$ & $95.88\%$ & $95.17\%$ & $95.20\%$ & $4.1\%$\\ 
   & $(\pm 0.46\%)$ & $(\pm 0.40\%)$ & $(\pm 0.20\%)$ & $(\pm 0.20\%)$ &\\ 
Malytics (Dex, MBR=0.2) & $\bf 05.27\%$ & $\bf 98.45\%$  &  $96.55\%$ & $98.65\%$ & $3.7\%$\\ 
 & $(\pm 0.63\%)$ & $(\pm 0.69\%)$ & $(\pm 0.55\%)$ & $(\pm 0.21\%)$ &\\
  \midrule
  Malytics (Zero-day) & $ \bf 10.59\%$ & $  \bf 96.31\%$ & $  \bf 92.68\%$ & $ 92.99\%$ & $ 3.4\%$\\ 
 \cite{mariconti_mamadroid} & $ 12.00\%$ & $86.00\%$ & $ 87.00\% $ &   $ - $ & $ - $\\ 
 \bottomrule
\end{tabular}
}
\caption{The Mean and Std of Malytics for DexShare dataset on the APK, Dex. Also, the results when the dataset is imbalanced and for zero-day (novel families) detection.}
\label{tab:mainTable3}
\end{table}

For real word application, we can increase the size of hash vector to improve the performance while Malytics still requires a light computation. Motivated from \cite{chen2018efficient}, we replaced the \emph{tf}-simashing weights (i.e -1 and 1 values) with a sparse matrix including -1, 1 and 0 \cite{li2006very}. We set the sparsity to $1\%$ and the size of \emph{tf}-simashing vector is 3000. Thus, only 30 elements of the hashing matrix are non-zero but after summing over the entire vocabulary, the hash size is 3000. Then, this vector is used as the input to the kernel layer and then the output layer. This sparse setting helps reduce the complexity of the \emph{tf}-simashing computation while increasing the hidden feature representation size. Table \ref{tab:sparse} presents the results of the experiment on both datasets. Malytics False positive improves slightly while hit-rate is very close to dense setting (see table \ref{tab:mainTable1}). The size of hashing provides richer hidden representation for DexShare samples. In addition to being more precise, Malytics has better hit-rate (see table \ref{tab:mainTable2}). The imbalanced setting shows Malytics performance for real word application. We set hash size 3000 which has slight impact on the LEM computation.

We conducted the last experiment to show Malytics can perform in different settings and its improvement capability. To have a comparable settings, other experiments of the paper are all based on dense matrix setting with the hash size 1024.

\begin{table}[ht]
\centering
\scalebox{0.8}{
\begin{tabular}{l*{5}{c}}
\toprule
 \bf Dataset  & \bf FNR &\bf Precision &\bf f1-score & \bf Accuracy & \bf FPR\\
 \midrule
Drebin & $01.53\%$ &  $96.68\%$ & $97.56\%$ & $97.54\%$ & $ 3.38\%$\\ 
      & $(\pm 0.50\%)$ & $(\pm 0.45\%)$ & $(\pm 0.40\%)$ & $(\pm 0.30\%)$ &\\ 
DexShare & $04.72\%$ & $96.69\%$ & $95.96\%$ & $96.00\%$ & $3.30\%$\\ 
   & $(\pm 0.50\%)$ &  $(\pm 0.35\%)$  &  $(\pm 0.38\%)$ & $(\pm 0.35\%)$ &\\ 
DexShare (MBR=0.2) & $\bf {04.42}\%$ & $98.91\%$ & $97.21\%$ & $98.90\%$ & $ \bf {2.60}\%$\\ 
   & $(\pm 0.27\%)$ &  $(\pm 0.82\%)$  &  $(\pm 0.53\%)$ & $(\pm 0.18\%)$ &\\ 
 \bottomrule
\end{tabular}
}
\caption{The Mean and Std of Malytics with sparse \emph{tf}-simashing for both Drebin and DexShare Datasets.}
\label{tab:sparse}
\end{table}

\subsection{Windows Malware Detection}

This section presents the capability of Malytics to detect Microsoft Windows malware. We show that the scheme is not restricted by any specific feature of the operating system. 

Table \ref{tab:mainTable4} shows that Malytics and Wuechner \cite{wuechner2017leveraging} outperform other models when it comes to distinguishing original Windows PE clean files from PE malware. Malytics is the most capable method in detecting malware with lowest the FPR compared with all methods, in particular, with our machine learning baselines. It has better FNR compared with Wuechner \cite{wuechner2017leveraging} and both schemes have the same precision while Wuechner \cite{wuechner2017leveraging} used an imbalanced dataset. In an imbalanced setting, trade-off indices are more reliable for concluding an analysis. F-score of the proposed model in Wuechner \cite{wuechner2017leveraging} is more than Malytics but the difference is not statistically significant. AUC indices show that Malytics outperforms other models. The FPR as an important factor for Windows platform malware analyzer is well less than $1\%$ that is critical for Windows. 

Table \ref{tab:mainTable5} shows that the proposed solution outperforms other models over all evaluation indices. It is to be expected that all models performance is reduced compared with table \ref{tab:mainTable4}, mainly because the benign set of this experiment is a mix of Windows PEs and third-party PEs while the malware set is from the same source and only has more samples. 

An interesting result of trying to distinguish Mal2016 malware set from WinAppPE benign set is in the comparison of DNN with Raff \cite{raff2017malware}. Raff \cite{raff2017malware} used deep CNN for detection. The results show deep learning models also can be competitive for malware application domain. Although the input of CNN is an image representation of the PE files and DNN's input is \emph{tf}-simhashing, we think deep learning models can also be competitive if the feature representation has more theoretical justification in the deep learning models' training algorithm. 

\begin{figure}[ht]
\centering
\includegraphics[scale=0.23]{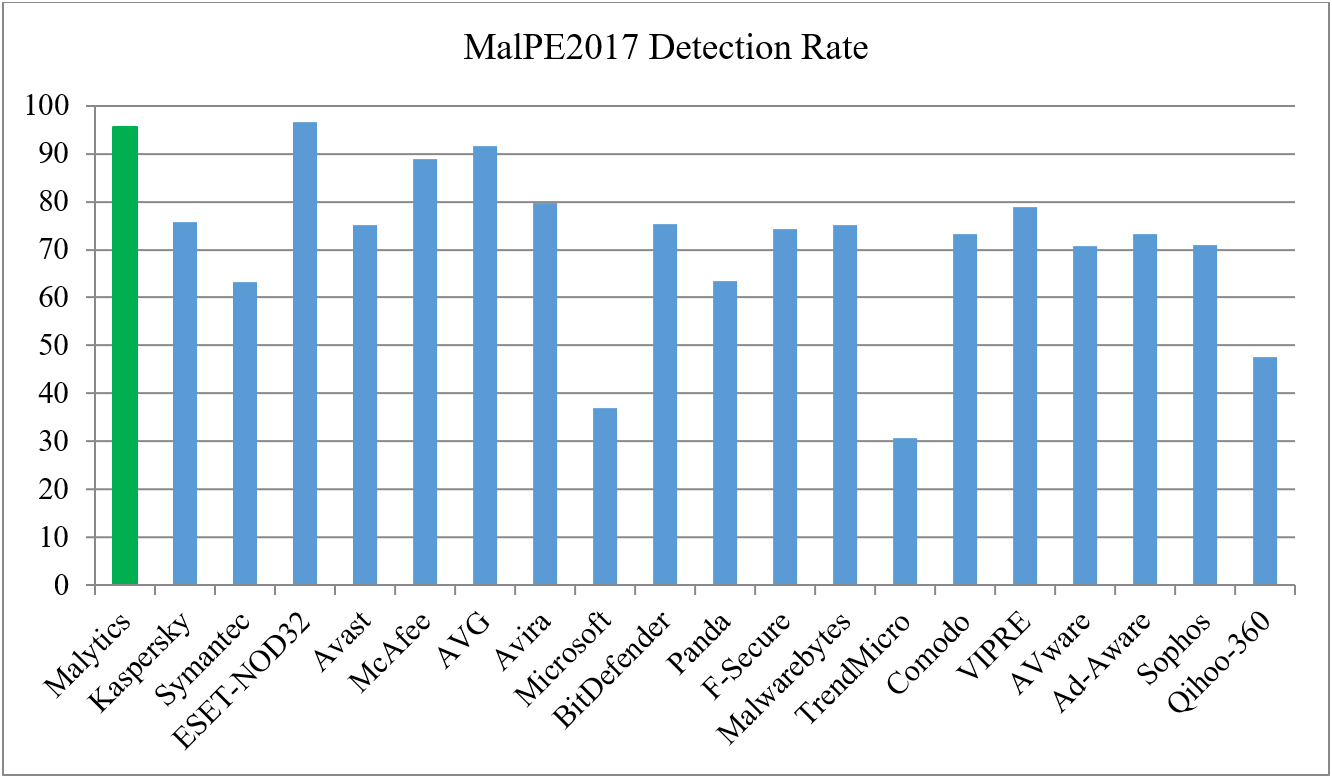}
\caption{The detection rate of Malytics and the 19 malware vendor for Mal2017, trained on Mal2016 and WinAppPE.}
\label{fig:FNR}
\end{figure}

Figure \ref{fig:FNR} shows the detection rate of Malytics on Mal2017. The training sets are Mal2016 and WinAppPE while Mal2017 is the test set. To have a similar setting to the real world, the training benign set was WinAppPE rather than WinPE. This experiment is to evaluate how well the scheme can detect zero-day attacks. We assume that a one year interval between the malware set in training and the malware set for testing is an acceptable chronological gap.

Malytics could successfully detect 95.5\% of the Mal2017 as zero-day samples. It is only one percent less than ESET-NOD32 detection rate. AVG with 91.5\% is the third in the ranking. Our ground truth for the detection rate of AV vendors is VirusTotal real-time update report. VirusTotal always uses the latest update of AVs; thus, the detection rates it reports are considerably better than they would be if the virus detector data was one year old. After one year, Malytics performs competitively with the best AV vendor software fully up to date. In another experiment, we trained Malytics using Mal2016 and WinPE sets and tested on Mal2017. As it is to be expected, the detection rate increases to 98.1\%.

\begin{table}[ht]
\centering
\scalebox{0.75}{
\begin{tabular}{l*{6}{c}}
\toprule
 \bf Model  & \bf FNR &\bf Precision &\bf f1-score & \bf Accuracy & \bf AUC & \bf FPR\\
 \midrule
SVM   & $1.30\%$ & $99.13\%$ & $98.91\%$ & $98.92\%$ &  $98.92\%$   & $0.86\%$\\ 
      & $(\pm 0.30\%)$ & $(\pm 0.25\%)$ & $(\pm 0.26\%)$ & $(\pm 0.26\%)$ &  $(\pm 0.25\%)$  &\\ 
XGBoost  & $1.30\%$ & $98.43\%$ & $98.56\%$ & $98.56\%$ &   $98.57\%$  & $1.57\%$\\ 
   & $(\pm 0.27\%)$ & $(\pm 0.26\%)$ & $(\pm 0.19\%)$ & $(\pm 0.19\%)$ &    $(\pm 0.19\%)$ &\\ 
DNN & $2.51\%$ & $96.77\%$ & $97.11\%$ & $97.09\%$ &  $97.09\%$   &$3.31\%$\\ 
      & $(\pm 0.59\%)$ & $(\pm 2.04\%)$ & $(\pm 0.78\%)$ & $(\pm 0.83\%)$ &   $(\pm 0.83\%)$  &\\ 
RF & $2.18\%$ & $98.44\%$ & $98.13\%$ & $98.14\%$ &  $98.14\%$   & $1.55\%$\\ 
 & $(\pm 0.35\%)$ & $(\pm 0.32\%)$ & $(\pm 0.22\%)$ & $(\pm 0.21\%)$ &    $(\pm 0.25\%)$   &\\
K-NN & $1.56\%$ & $98.50\%$ & $98.47\%$ & $98.47\%$ &  $98.47\%$   &  $1.50\%$\\ 
      & $(\pm 0.38\%)$ & $(\pm 0.04\%)$ & $(\pm 0.20\%)$ & $(\pm 0.20\%)$ &  $(\pm 0.20\%)$      &\\ 
ELM & $ 1.00\%$ & $ 95.82\%$ & $ 97.38\%$ & $ 97.34\%$ &  $97.79\%$  & $ 4.30\%$\\ 
 & $(\pm 0.17\%)$ & $(\pm 0.35\%)$ & $(\pm 0.18\%)$ & $(\pm 0.19\%)$ &      $(\pm 0.18\%)$ &\\
Malytics & $ \bf 0.55\%$ & $ \bf 99.20\%$ & $ 99.32\%$ & $ \bf 99.32\%$ &      $ \bf 99.96\%$  & $\bf 0.80\%$\\ 
 & $(\pm 0.23\%)$ & $(\pm 0.27\%)$ & $(\pm 0.12\%)$ & $(\pm 0.11\%)$ &    $(\pm 0.19\%)$   &\\
  \midrule
 \cite{wuechner2017leveraging} & $ 1.00\% $ & $ \bf 99.20\% $ & $ \bf \bf 99.70\% $ & $-$ &  $99.30\%$  & $ - $\\
  & $(\pm 0.00\%)$ & $(\pm 0.00\%)$ & $(\pm 0.5\%)$ & $-$  &    $(\pm 0.1\%)$   &\\
 \cite{carlin2017dynamic} & $ 0.80\% $ & $ - $ & $ 99.10\% $ & $99.05\%$ &         & $ 1.10\% $\\ 
 \bottomrule
\end{tabular}
}
\caption{The Mean and Std of Malytics and the baselines for WinPE and Mal2016 of the PEShare Dataset.}
\label{tab:mainTable4}
\end{table}

\begin{table}[ht]
\centering
\scalebox{0.75}{
\begin{tabular}{l*{6}{c}}
\toprule
 \bf Model  & \bf FNR &\bf Precision &\bf f1-score & \bf Accuracy & \bf AUC & \bf FPR\\
 \midrule
SVM   & $2.78\%$ & $98.32\%$ & $97.76\%$ & $97.78\%$ &  $97.78\%$   & $1.7\%$\\ 
      & $(\pm 0.13\%)$ & $(\pm 0.16\%)$ & $(\pm 0.14\%)$ & $(\pm 0.14\%)$ &  $(\pm 0.14\%)$  &\\ 
XGBoost  & $2.24\%$ & $97.61\%$  &  $97.68\%$ & $97.68\%$ &  $97.69\%$  & $2.4\%$\\ 
 & $(\pm 0.36\%)$ & $(\pm 0.19\%)$ & $(\pm 0.20\%)$ & $(\pm 0.20\%)$&    $(\pm 0.20\%)$   &\\ 
DNN & $4.43\%$ & $94.71\%$ & $95.09\%$ & $95.02\%$ &  $95.02\%$   & $5.5\%$\\ 
 & $(\pm 0.60\%)$ & $(\pm 3.90\%)$ & $(\pm 1.90\%)$ & $(\pm 2.00\%)$ &     $(\pm 2.05\%)$   &\\
RF & $4.46\%$ & $97.85\%$  &  $96.68\%$ & $96.72\%$ &   $96.72\%$   & $2.1\%$\\ 
 & $(\pm 0.58\%)$ & $(\pm 0.26\%)$ & $(\pm 0.29\%)$ & $(\pm 0.27\%)$&     $(\pm 0.28\%)$   &\\ 
K-NN & $2.30\%$ & $96.90\%$ & $97.30\%$ & $97.29\%$ &   $97.29\%$    & $3.1\%$\\ 
 & $(\pm 0.15\%)$ & $(\pm 0.31\%)$ & $(\pm 0.19\%)$ & $(\pm 0.19\%)$ &    $(\pm 0.19\%)$  &\\
ELM & $ 2.18\%$ & $ 93.95\%$ & $ 95.84\%$ & $ 95.76\%$ &  $97.42\%$   & $ 6.3\%$\\ 
 & $(\pm 0.17\%)$ & $(\pm 0.42\%)$ & $(\pm 0.19\%)$ & $(\pm 0.14\%)$ &      $(\pm 0.19\%)$  &\\
Malytics & $ \bf 1.32\%$ & $ \bf 98.65\%$ & $\bf 98.66\%$ & $ \bf 98.67\%$ &      $ \bf 99.81\%$  & $\bf 1.3\%$\\ 
 & $(\pm 0.06\%)$ & $(\pm 0.04\%)$ & $(\pm 0.20\%)$ & $(\pm 0.19\%)$ &    $(\pm 0.22\%)$   &\\ 
  \cite{raff2017malware} & $ - $ & $ \bf - $ & $ \bf - $ & $94.00\%$ &  $98.10\%$  & $ - $\\
 \bottomrule
\end{tabular}
}
\caption{The Mean and Std of Malytics and the baselines for WinAppPE and Mal2016 of the PEShare Dataset.}
\label{tab:mainTable5}
\end{table}

\subsection{\emph{tf}-simhashing visualization}\label{TFsimhashingVisualization}

To have more insight into the proposed latent representation, that is, \emph{tf}-simhashing layer and RBF kernel, a visualization experiment was conducted on the feature space. We implemented the experiment on Windows malware dataset, WinPE and Mal2017 and used t-SNE \cite{maaten2008visualizing} to visualize the space.

To this end, \emph{tf}-simhashing values have been clustered using k-means clustering technique and then the centroids of the clusters fed to t-SNE. We think because in the k-means optimization algorithm the Euclidean distance is used as a metric, it can provide a similar ground with RBF kernel that is also based on Euclidean distance. But RBF kernel provides an infinite feature space that we cannot visualize easily. Also, the intention of the experiment is to show that similar vectors of \emph{tf}-simhashing are quantified similarly and \emph{tf}-simhashing is meaningful. 

More precisely, \emph{tf}-simhashing of the dataset (here 17824 = 2$\times$8912) benign and malware samples) are clustered into 2400 clusters (1200 centroids per class). So, the input of the k-means function is a matrix of size 17824$\times$1024 and the output is a matrix of size 2400$\times$1024. Effectively, we use k-means as a vector quantisation algorithm, yielding on average, one codebook per about 7.4 vectors ($\frac{17824}{2400}=7.43$). Our experiment shows the 2400-vector representation provides a good visual picture to understand the dataset. The matrix of 2400$\times$1200 is fed to t-SNE to be mapped into two-dimensional space (2400$\times$2) for visualization.

\begin{figure}[ht]
\centering
\includegraphics[scale=0.60]{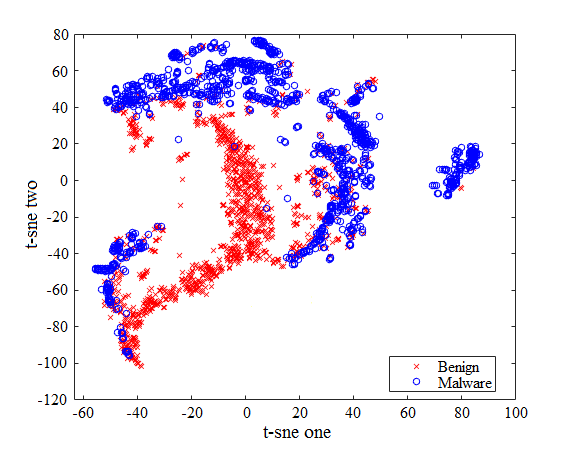}
\caption{t-SNE of WinPE versus Mal2016. WE clustered each class with 1200 centroids for each class.}
\label{fig:test}
\end{figure}

We obtain codebooks from malware and benign sets separately. That is, 1200 centroid per class. Figure \ref{fig:test} shows clustering the representation provides a meaningful result and the codebooks are basically distributed similarly. This means the vectors of \emph{tf}-simhashing is meaningful and can settle in closed distance when we optimize k-mean with its Euclidean distance measure. We do not show representation of feeding \emph{tf}-simhashing to t-SNE directly because it generates a meaningless distribution.

\subsection{Case study and Further analysis of the scheme}\label{FurtherAnalysisAndCaseStudy}

\subsubsection{Android family detection case}\label{AndroidFamilyDetectionCase}

The scope of this paper is not a particular malware family but covers the range of malware disseminating all over the network. Our datasets were collected with this purpose. However, looking into some specific cases may provide better inside into the model. We used the DexShare dataset for detail study.

As described in section \ref{tab:mainTable3}, we tested the model on 20 new families. In brief, we chose groups of 4 families for testing and the remaining 16 families plus all the other malware and benign sets for training. This routine was repeated 5 times. 

\begin{figure}[ht]
\centering
\includegraphics[scale=0.23]{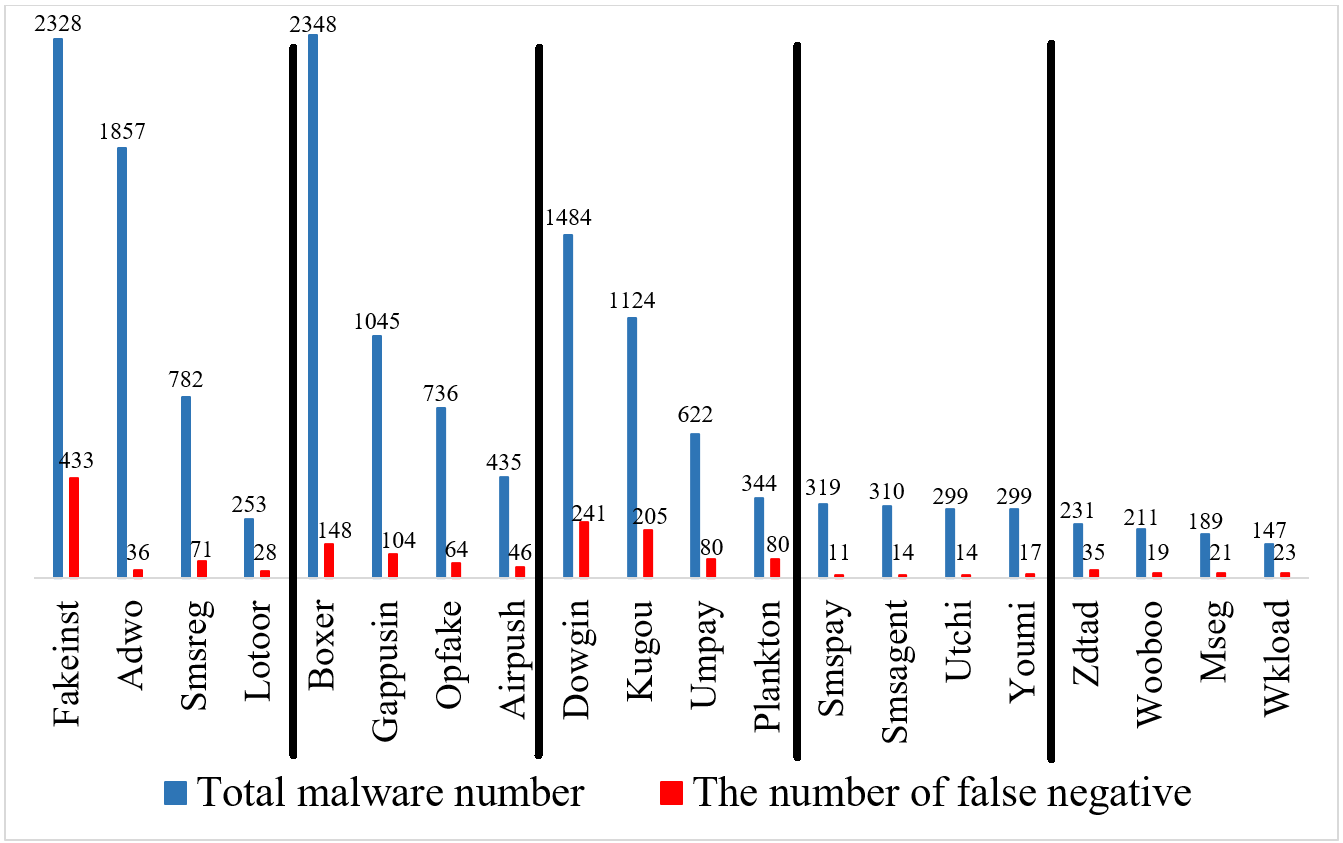}
\caption{Total number of samples and False negative detection based on 4 families fetching out on DexShare. The black bar separates each group of 4 families.}
\label{fig:FamiliesAndroid}
\end{figure}

Figure \ref{fig:FamiliesAndroid} presents the number of  total samples in each family and false negative detection. We tested Malytics in the balanced and imbalanced group of families. Also, one of the groups consists of malware families (Fakeinst\footnote{{https://www.f-secure.com/v-descs/trojan\_android\_fakeinst.shtml}}, Adwo\footnote{{https://www.sophos.com/en-us/threat-center/threat-analyses/adware-and-puas/Android\%20Adwo/detailed-analysis.aspx}}, SMSreg\footnote{{https://home.mcafee.com/virusinfo/virusprofile.aspx?key=8503749}}, Lotoor\footnote{{https://www.symantec.com/security\_response/writeup.jsp?docid=2012-091922-4449-99}}) with 4 different functions/intention. The detection rate is similar for most families and the diversity of malware function did not prevent detection.

The Adwo family is the least challenging for detection by Malytics. This is to be expected since although Adwo is not in the training set, other adware variants are used in training. Fakeinst and Opfake were reported as similar families and Fakeinst had been continued to be detected while Opfake not \footnote{{https://threatpost.com/opfake-fakeinst-android-malware-variants-continue-resist-detection-080712/76887/}}. Malytics could detect Opfake better than Fakeinst. It might because of the complexity of the Fakeint that our model could not detect it well, as it was also continued to disseminate over the net in the real world. But, it might because of the number of sample in training when another family is presented only in testing set. 

The worst detection rate belongs to Plankton\footnote{{https://www.f-secure.com/v-descs/trojan\_android\_plankton.shtml}} ($\frac{80}{344}=23.3\%$). This family sits silently, collecting information and sending it to a remote location. Its variants have a wide range of actions\footnote{{https://www.avira.com/en/support-threats-summary/tid/8996/threat/ANDROID.Plankton.C.Gen}}. Calleja \cite{calleja2018picking} has particular analysed on Plankton. They showed that this family is very similar to the Nyleak and BaseBridge families. These two similar families to Plankton have only 28 and 2 samples in DexShare. We think that because, in training, there are few similar malware samples to the Plankton family, Malytics' detection rate is reduced for this family; however, the 76.7\% hit rate is still very good for this family in this setting.

\begin{figure}[ht]
\centering
\includegraphics[scale=0.43]{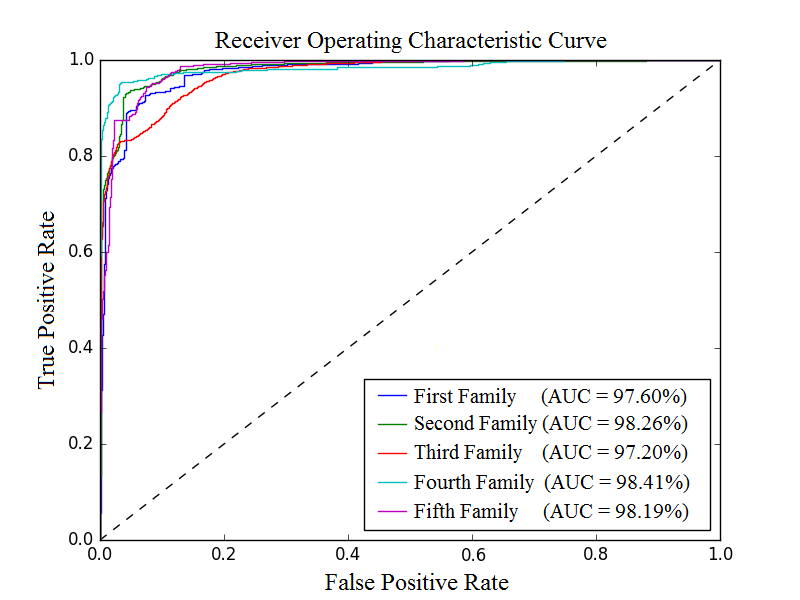}
\caption{The ROC curve for novelty detection on DexShare. Each curve shows the ROC curve and AUC when groups 4 families have been fetched out from DexShare. First family is the first left 4 family in the figure \ref{fig:FamiliesAndroid} and so on.}
\label{fig:FamiliesROC}
\end{figure}

Figure \ref{fig:FamiliesROC} presents ROC and respective AUC of the novelty detection. With 2-3\% FPR, the hit rate is more than 75\% for all four families that seems promising.

\subsubsection{Random kernel}\label{RandomKernel}

Malytics is based on the batch learning algorithm. That is, the entire training set is fed to Malytics for training in one batch. This process might become computationally very expensive and demand a large amount of memory for big data.

\begin{figure}[ht]
\centering
\includegraphics[scale=0.21]{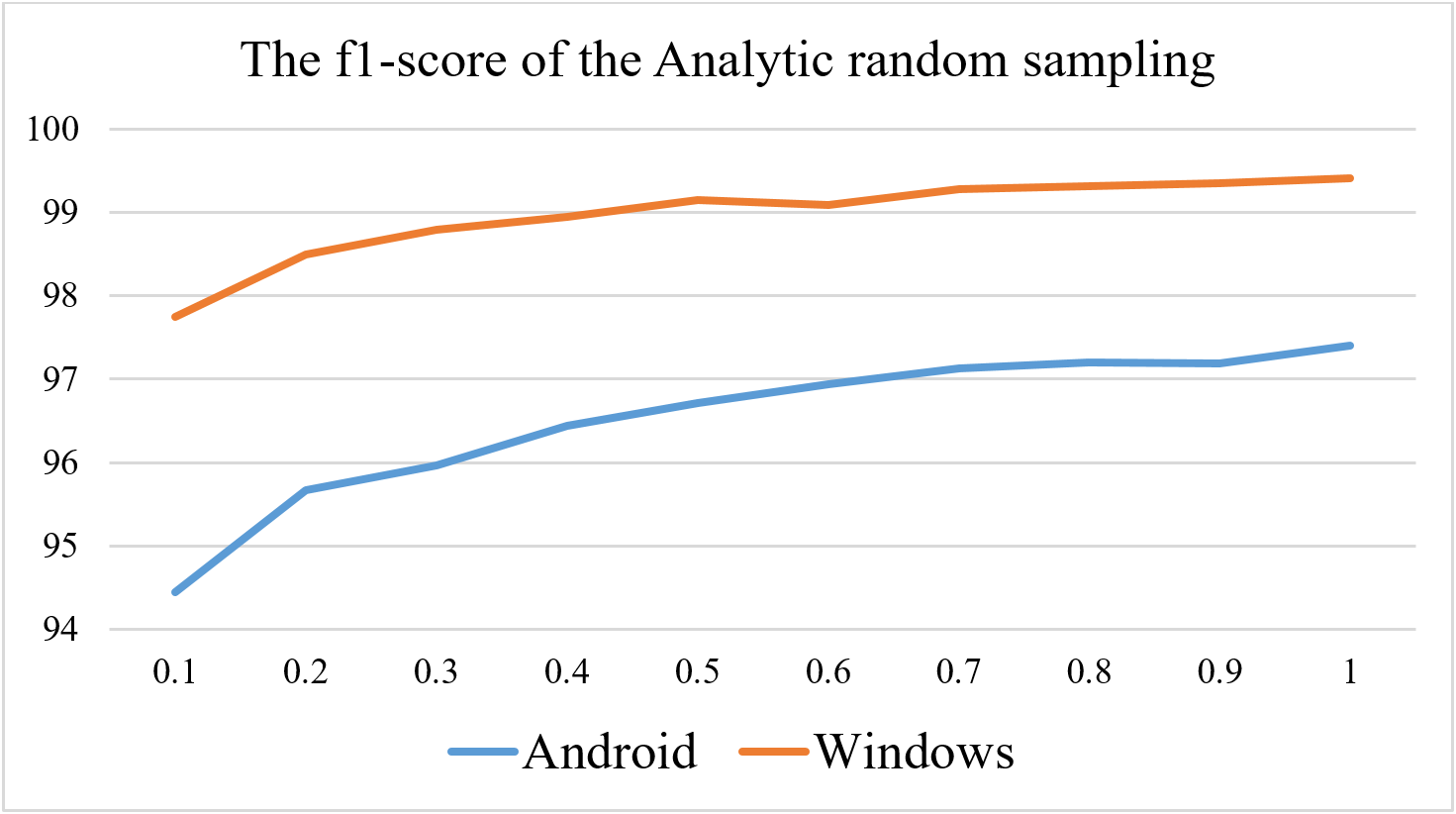}
\caption{The f1-score of Malytics trained on a random subsets of the datasets. For Android, Drebin dataset was used and for Windows, WinPE versus Mal2016.}
\label{fig:Randomkernel}
\end{figure}

As mentioned in \ref{Scheme}, Malytics has the capability of being trained on a random subset of training data while still keeping the generalization performance. This random selection has a negative impact on the performance of Malytics. Figure \ref{fig:Randomkernel} presents the f-score for both platforms when the kernel matrix size varies from 10\% to 100\% of the original kernel matrix. Similar to all previous evaluations, Android platform is more impacted than Windows. It is to be expected because Windows PEs are all from Microsoft while Android apps are developed by many different developers and therefore exhibits a greater variety. 

Table \ref{tab:runtimeRandom} shows the run-time performance of Malytics with random kernel sampling from 10\% to 100\% of the kernel matrix. The training and testing time increases but not sharply. In contrast, the f-score value increases sharply initially (see figure \ref{fig:Randomkernel}). This shows that choosing more than a threshold number of samples may provide the desired performance with optimum memory and computation requirements. This demonstrates Malytics's scalability.

\begin{table*}[ht]
\centering
\begin{tabular}{l*{11}{c}}
\toprule
 \bf Dataset & Phase & \bf $L=\frac{N}{10}$ &\bf $L=\frac{N}{20}$ & \bf $L=\frac{N}{30}$ &\bf $L=\frac{N}{40}$ &\bf $L=\frac{N}{50}$ & \bf $L=\frac{N}{60}$ &\bf $L=\frac{N}{70}$ & \bf $L=\frac{N}{80}$ &\bf $L=\frac{N}{90}$ & \bf $L=N$\\
 \midrule
 \bf Drebin & Train & 0.12s & 0.15s & 0.29s & 0.58s & 0.80s & 1.20s & 1.75s & 2.35s & 3.20s & 3.60s\\  
 & Test & 0.10s & 0.16s & 0.22s & 0.27s & 0.35s & 0.40s & 0.45s & 0.54s & 0.58s & 0.6s \\ 
 \midrule
 \bf WinPE vs Mal2016 & Train & 0.09s & 0.35s & 0.78s & 1.50s & 2.40s & 3.70s & 5.30s & 7.20s & 9.60s & 11.6s\\  
 & Test & 0.19s & 0.34s & 0.52s & 0.70s & 0.80s & 1.14s & 1.20s & 1.30s & 1.50s & 1.55s\\ 
 \bottomrule
\end{tabular}
\caption{The run-time performance of Malytics with randomly selected kernel matrix $L$ ranging from 10\% to 100\% of the kernel matrix $N$. The time is for each fold (i.e. 8888 training and 2222 test samples) of the 5-fold cross-validation.}
\label{tab:runtimeRandom}
\end{table*}

\subsubsection{Run-time performance}

Because Malytics is based on static analysis, we compare it with MAMADROID \cite{mariconti_mamadroid}. The run-time performance of Malytics is not dependent on the operating system while MAMADROID is proposed to detect Android malware. 

The average execution time of MAMADROID for benign samples in family and package modes are 27.3s and 33.83s per sample respectively. MAMADROID says for 10000 apps that are being submitted to Google Play per day, the model requires less than one and a half hours to complete execution with 64 cores.

In our prototype of Malytics, on average, \emph{tf}-simhashing algorithm speed is 560KB per second. This includes extracting a dex file out of the APK. Also, for the Drebin dataset with 11110 samples (on average $\frac{3+0.36}{2}=1.68MB$), the training and testing time are 3.6s and 0.6s respectively, for each fold (from table \ref{tab:runtimeRandom}). That is, $\frac{3.6s}{11110\times0.8}=0.4ms$ per sample to train the model and $\frac{0.6s}{11110\times0.2}=0.27ms$ per sample to test an App.

Based on the statistics of Drebin malware set and random benign set collected from Google Play, Malytics requires, approximately, $\frac{1.68MB}{0.560MB/Sec}=3$ second for each sample to hash and 0.27ms to detect. For 10000 apps, 30003 seconds to complete execution with one core and 470 seconds with 64 cores. In short, Malytics needs less than 8 minutes to complete execution for all the Google Play new samples in a day. The speed of Malytics provides the possibility of frequently training with new samples.

\section{Limitations and future work}

The main limitation of Malytics is the amount of the required memory. More precisely, the advantage of Malytics batch learning technique is speed and mostly convexity; nevertheless, the model needs to store all input samples as a batch in memory. The learning process also requires a considerable amount of memory to obtain the output layer weights, of course, but only for big data. Also, for the test phase, the kernel layer needs to keep all the training samples. This intensifies the memory issue and also makes the test process directly dependent on the training set.

Malytics hashes the whole binary file into one vector. This may not be effective for binary files that are partly infected with malicious codes. To address this issue, hashing windows of a binary file may help and is a potential direction of future study.

Malytics relies on the static analysis of binary files and it does not have any knowledge of the behavior of the binary. Despite the computational expenses of dynamic analysis, it can cover deficiencies of the static features of malware in particular for obfuscated malware and more importantly for advanced persistent attacks. 

\emph{tf}-simhashing visualization experiment suggests that k-means might be used for feature learning the hashing dictionary rather than random generation from a distribution. Also, deep learning models have no theoretic underlying with a random initialization layer. This may contribute to deep learning models as well.

The dex file is the not the only source of information in Android apps. The manifest file also contains critical information. For Windows, PEs structure can be informative if we could embed this information into the hashing algorithm. Also, developing a feasible algorithm for larger n-grams may improve the performance. Finally, an ensemble of fast learning models is a potential direction of study for future work.

\section{Related Work}

In addition to malware as a general concept, there are novel systems to deal with particular types of malware \cite{kharraz2016unveil,AndroidRootExploits}. In both cases, learning-based systems show very promising results \cite{saracino2016madam,ucci2017survey,egele2012survey,8101555}.

For Android malware detection, Mariconti \cite{mariconti_mamadroid} proposed a static-feature extraction model that could provide a very good performance. A novelty of the work was the proposed random variable based on Markov chain. The output was fed to a feature extraction phase in which Principal Component Analysis (PCA) \cite{wold1987principal} was used. At the end, each sample file yeilds a vector of size 100,000 to be classified as either benign or malware. Applying PCA to such feature space requires a huge amount of memory to obtain the co-variance matrix. Zhu \cite{zhu2017droiddet} recently showed that rotational forest, as a classifier, has the capability of being applying for Android malware detection. They also used static features.

Wuechner \cite{wuechner2017leveraging} used a compression-based graph mining technique to detect Windows malware. They widely evaluated the effect of classifiers on their scheme and reported that all the applied classifiers provide similar results. Carlin \cite{carlin2017dynamic} used the run-time opcodes of every sample with significantly different approach compared with Wuechner \cite{wuechner2017leveraging} and still presented competitive results. Both papers used dynamic analysis of Windows PEs. Dynamic analysis of malware is more computationally expensive than static analysis. Having said that, it is well-known to be used in many anti-malware production due to its reliable performance and capability to cope with obfuscated files. 

Dynamic analysis of a malware does not influence the vulnerability of machine learning because the feature space can be still reverse engineered to craft an adversary. Very recently, Stokes et al. \cite{stokes2017attack} proposed a detection system using dynamic analysis and they showed it is still vulnerability to crafted adversarial attacks.

The \emph{tf}-simhashing representation is a promising feature representation \cite{yousefi2017fast}. Simhashing has also been used for malware detection and detecting similarity between data/files \cite{han2013malware,sadowski2007simhash}. In both \cite{han2013malware,yousefi2017fast}, simhashing was used to represent each file as an image which was then fed to a naive classifier or a CNN. Malware detection on the basis of visualization is not restricted only to simhashing. Raff \cite{raff2017malware} proposed a new feature representation to map any binary file into an image and used CNN as the classifier.

ELM shows very promising results for malware activity detection and identification of malicious packed executables. Kozi \cite{kozik2018distributing} presented a distributed ELM using NetFlow data structure alongside the Apache Spark \footnote{\url{https://spark.apache.org/}} framework that provided good performance. Different types of ELM have been applied for malicious packed executable identification \cite{xie2016absent}.

\section{Conclusion}

In this paper, we proposed a learning-based malware detection model called "Malytics". This integrated model comprises two layers of latent feature representation and a layer for prediction. The first layer is a hashing algorithm (\emph{tf}-simhashing) and we showed that it has a close relation to the first layer of the Extreme Learning Machine (ELM). ELM is the output layer of the proposed scheme. We showed that having a layer to measure the similarity of \emph{tf}-simhashing before output layer strongly improves the performance of the scheme. We used the RBF kernel for the similarity measure. 

We conducted comprehensive evaluations on Drebin, DexShare, PEShare datasets and Malytics outperforms different baselines including non-ensemble state-of-the-art models. Drebin and DexShare are Android apps and PEShare is Windnows PEs. The dex file of Android apps is informative enough to compete with related work. We tested how well Malytics could perform on imbalanced datasets, for novel family detection. Novelty detection was organized in two different ways: particular family detection and chronological novelty detection. We also evaluated the speed and scalability of Malytics. It shows promising results for large-scale data.

\begin{appendices}


\section{Optimization}\label{appendix:B}

The dual problem can be optimized as follows:

\begin{equation}\label{1}
\begin{split}
\frac{\partial L_{\text{Dual}_{\text{ELM}}}}{\partial \boldsymbol  \beta_j}=0 \to \beta_j=\displaystyle\sum_{i=1}^N\alpha_{i,j}h(x_i)^T \to \boldsymbol\beta= \bf H^T \boldsymbol\alpha
\end{split}
\end{equation}

\begin{equation}\label{2}
\frac{\partial L_{\text{Dual}_{\text{ELM}}}}{\partial\xi_j}=0 \to \boldsymbol \alpha_i=C\xi_i, i = 1, ..., N
\end{equation}

\begin{equation}\label{3}
\begin{aligned}
\frac{\partial L_{\text{Dual}_{\text{ELM}}}}{\partial \boldsymbol \alpha_j}=0 \to \boldsymbol h(x_i) \boldsymbol \beta- {\bf t}_{i}^{T}+\xi_{i}^{T}=0, i = 1, ..., N
\end{aligned}
\end{equation}

Where $\boldsymbol\alpha_i=[\alpha_{i,1}, ..., \alpha_{i,m}]^T$ and $\boldsymbol\alpha_i=[\boldsymbol\alpha_{1}, ..., \boldsymbol\alpha_{N}]^T$. With a bit of calculus, for $\boldsymbol\beta$:

\begin{equation}
\begin{split}
\boldsymbol\beta={\bf H}^T \big(\frac{\bf{I}}{C}+\bf{H}\bf{H}^T)^{-1}\bf{T}
\end{split}
\end{equation}

\end{appendices}

\footnotesize{
\bibliographystyle{IEEEtran}
\bibliography{References} \label{References1}
}

\end{document}